\documentclass[prl,twocolumn,showpacs,amsmath]{revtex4}
\usepackage{graphicx, bm}
\usepackage[dvips]{color}

\begin{document}
\title{Inelastic X-ray Scattering Study of SmFeAs(O$_{1-x}$F$_{y}$) Single Crystals:
Evidence for Strong Momentum-Dependent Doping-Induced Renormalizations of Optical Phonons}

\author{M. Le Tacon$^{1,2}$, T. R. Forrest$^{3}$, Ch. R\"{u}egg$^{3}$, A. Bosak$^{1}$, A. C. Walters$^{1,3}$, R. Mittal$^{4}$, H. M. R\o nnow$^{5}$, N. D. Zhigadlo$^{6}$, S. Katrych$^{6}$, J. Karpinski$^{6}$, J. P. Hill$^{7}$, M. Krisch$^{1}$, D.F. McMorrow$^{3}$}
\affiliation{$^{1}$ European Synchrotron Radiation Facility; BP 220; F-38043 Grenoble Cedex; France\\
$^{2}$Max-Planck-Institut f\"{u}r Festk\"{o}rperforschung; Heisenbergstra{\ss}e 1; D-70569 Stuttgart, Germany\\
$^{3}$London Centre for Nanotechnology and Department of Physics and Astronomy; University College London; London WC1E 6BT; United Kingdom\\
$^{4}$Solid State Physics Division; Bhabha Atomic Research Centre; Trombay; Mumbai 400 085; India\\
$^{5}$Laboratory for Quantum Magnetism; Ecole Polytechnique F\'ed\'erale de Lausanne; CH--1015 Lausanne; Switzerland\\
$^{6}$Laboratory for Solid State Physics ETH Z\"{u}rich; CH-8093 Z\"{u}rich; Switzerland\\
$^{7}$Department of Condensed Matter Physics and Materials Science; Brookhaven National Laboratory; Upton; New York 11973, USA\\
}

\date{\today}

\begin{abstract}
We report inelastic x-ray scattering experiments on the lattice dynamics in SmFeAsO and superconducting SmFeAsO$_{0.6}$F$_{0.35}$ single crystals. Particular attention was paid to the dispersions along the [100]
direction of three optical modes close to 23 meV, polarized out of the FeAs planes. Remarkably, two of these modes are strongly renormalized upon fluorine doping. These results provide significant insight into
the energy and momentum dependence of the coupling of the lattice to the electron system and underline the importance of spin-phonon coupling in the superconducting iron-pnictides.

\end{abstract}


\maketitle
\date{\today}

\par
The recent discovery of superconductivity in FeAs based compounds has stimulated considerable interest~\cite{Takahashi_Nature2008}. To date three distinct families of FeAs based superconductors with respective formulae XFeAs (X= Li or Na), MFe$_2$As$_2$ (M= Ca, Ba, Sr), REFeAsO (where RE is a rare earth) $-$ the so-called ``111"s, ``122"s and ``1111"s $-$ have been identified. Among them, the last family has the highest values of the superconducting transition temperature $T_c$: depending on the lanthanide, $T_c$ can be as high as 55~K~\cite{Ren_CPL2008}. The high value of $T_c$ and early DFT calculations~\cite{Boeri} initially appeared to support the notion of an exotic mechanism for superconductivity in these materials.
However, recent experiments have established that the role of phonons in the electron coupling cannot be
totally ignored. In particular, the observation of large Fe isotope effects~\cite{Liu_Nature2009}, the fact that the Fe-As configuration is intimately related to $T_c$~\cite{Mukuda_JPSJ2008} and to the magnetic state of the iron sublattice~\cite{Yin_PRL2008, Yildirim_PhysicaC2009} show that the lattice is actively involved in this problem. In addition, substantial deviations have been found between the experimentally observed phonon density of states (PDOS) and \emph{ab-initio} calculations~\cite{Fukuda_JPCS2008}, suggesting that these calculations do not captured all the relevant physics. A spin-polarized calculation is, for example, required to obtain good agreement with the experimental phonon dispersion in the case of undoped CaFe$_2$As$_2$, even above the magnetic ordering temperature $T_N$~\cite{Hahn_PRB2009}. Furthermore, it was recently demonstrated that zone center Raman active phonons are strongly renormalized at $T_N$ in Ba$_{1-x}$K$_x$Fe$_2$As$_2$~\cite{Rahlenbeck} and Ba(Fe$_{1-x}$Co$_x$)$_2$As$_2$~\cite{Chauviere_PRB2009}. These results all highlight the role of spin-phonon coupling in iron-pnictides.

In previous experiments on NdFeAsO$_{1-x}$F$_x$ powders, an unexpected signature of the doping in the PDOS was discovered~\cite{LeTacon_PRB2008}.
In particular, it was found that even at room temperature, at least one of the three phonon branches located around 23 meV softens on doping.
All the three modes are polarized out of the (Fe,As) planes, parallel to the c-axis, and involve motions of either the As or Fe ions. At least one of the modes couples directly to the magnetic iron sublattice~\cite{Yndurain_PRB2009, Yildirim_PhysicaC2009}.

Here, we utilize inelastic x-ray scattering (IXS) to investigate the phonon dispersions of SmFeAsO and superconducting SmFeAsO$_{0.6}$F$_{0.35}$ single crystals, with particular emphasis on these three modes. We find an unexpectedly strong doping-induced renormalization of the 21 meV (As,Sm) and 26 meV (Fe,O) modes. In addition, this renormalization displays a non-trivial momentum dependence. Taken together, these results provide significant insight into the evolution with doping of the energy and momentum dependence of the coupling between the lattice and the spin degrees of freedom, and suggest that the spin-phonon coupling is of key importance.

\par
High quality single crystals SmFeAsO and SmFeAsO$_{0.60}$F$_{0.35}$ were grown as described elsewhere~\cite{Zhigadlo_JPCM2008, Karpinski_PhysicaC2009}.
Our SQUID measurements on SmFeAsO  show magnetic ordering at $T_N =$ 130~K, and the superconducting transition in the nominally doped SmFeAsO$_{0.60}$F$_{0.35}$ takes place at $T_c =$ 53~K.
Contrary to 122 pnictide single crystals, the samples available are of very small size (about 100$\times$100$\times$20 $\mu $m$^3$ for the ones we used in this experiment) and consequently their phonon dispersions can only be investigated using IXS.
The IXS experiments were carried out on beamline ID28 at the European Synchrotron Radiation Facility, with a photon energy of 21.747 keV and an energy resolution of 2.0 meV. The x-ray beam was focused down to 50$\times$40 $\mu m^2$.
Rocking curves of 0.5$^\circ$ were measured on the (0 0 4) reflection, indicating a low c-axis mosaicity.
Phonon intensities are proportional to ($\textbf{Q}\cdot\bm{\sigma}_n$)$^2$~\cite{Burkel_RPP99}, $\bm{\sigma}_n$ being the eigenvectors of the modes and $\textbf{Q}=(Q_a~Q_b~Q_c)$ the total momentum transfer, defined here with respect to the tetragonal unit cell (a=b=3.936{~\AA}, c=8.485{~\AA} for SmFeAsO and a=b=3.931{~\AA}, c=8.472{~\AA} for SmFeAsO$_{0.60}$F$_{0.35}$). In what follows we also use  $\textbf{q}=(q_a~q_b~q_c)$ for the reduced momentum transfer. We therefore carried out our measurements in zones with the largest $Q_c$ component possible to maximize the intensity for the three c-axis polarized phonons we were interested in.
All results were obtained at room temperature.

\begin{figure}
\begin{center}
\includegraphics[width=0.95\columnwidth]{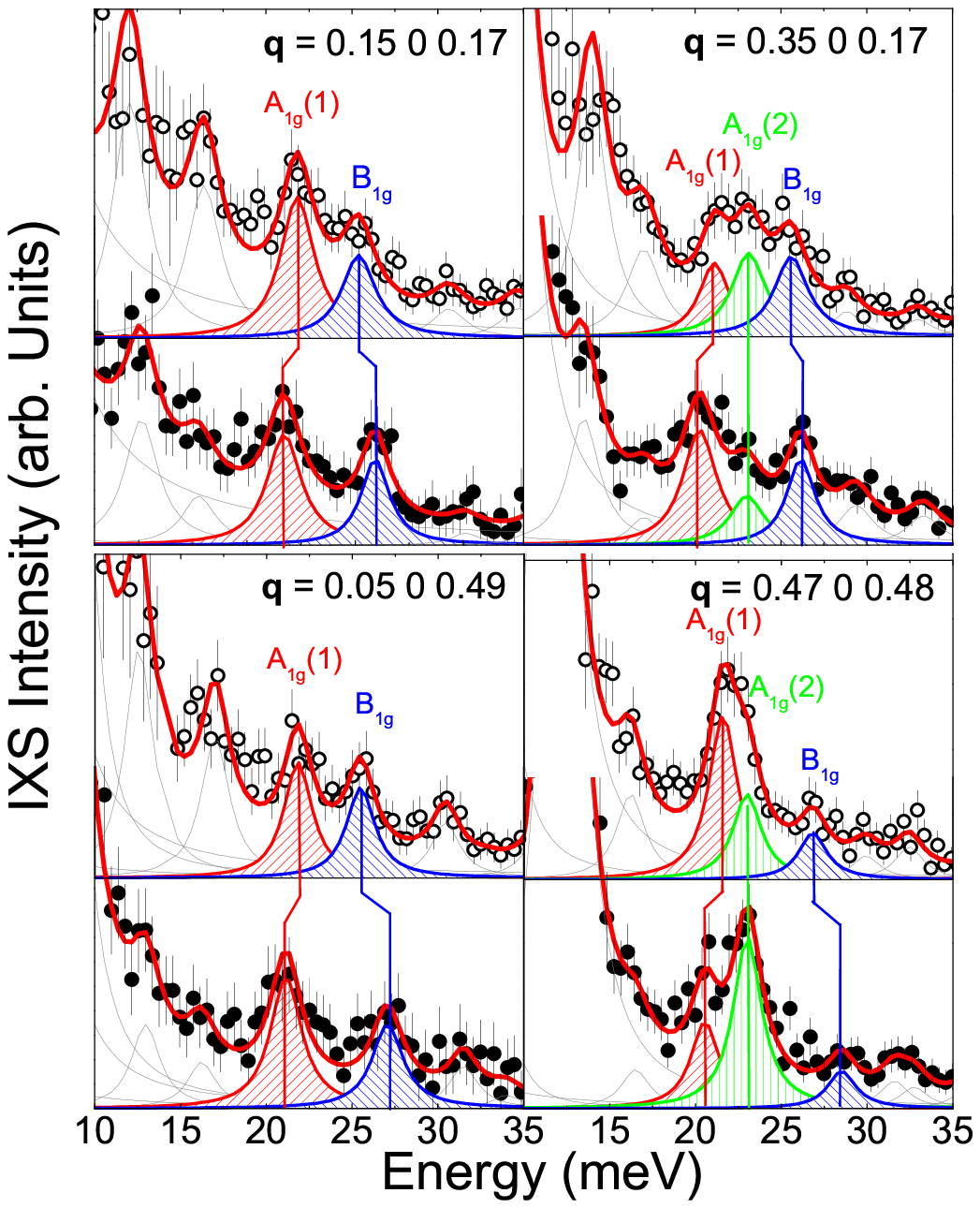}
\end{center}\vspace{-5mm}
\caption{(Color online) Doping dependance of the IXS spectra of SmFeAs(O$_{1-x}$F$_{y}$) for four \textbf{Q} positions [open circles: SmFeAsO, full circles: SmFeAsO$_{0.60}$F$_{0.35}$]. The red line is the result of the total fit, and the thin gray lines are the individual phonons. The three phonon lines around 23 meV have been highlighted. Clear energy shifts, with opposite signs, are observed in the 21 meV and 26 meV modes.}
\label{figdoping}
\end{figure}

Fig.~\ref{figdoping} shows the main results of our study. Here, phonon spectra for SmFeAsO and SmFeAsO$_{0.6}$F$_{0.35}$ are presented close to the (1 0 12) $\Gamma$ point at four different momentum transfers, corresponding to ($q_a$ 0 0.17) (upper panel) and approximately ($q_a$ 0 0.5) (lower panel).
Three features are observed around 23 meV. Two of them display a clear doping dependence.
The phonon located around 21 meV systematically softens on going from the undoped to the doped compound, while the phonon close to 26 meV hardens. The third feature, located around 23 meV, is only observed at large $q_a$ and shows essentially no doping dependence, except at \textbf{q} = (0.45 0 0.17) where it is softened by about 0.5 meV (not shown here).

\begin{figure}[b!]
\begin{center}
\includegraphics[width=0.95\columnwidth]{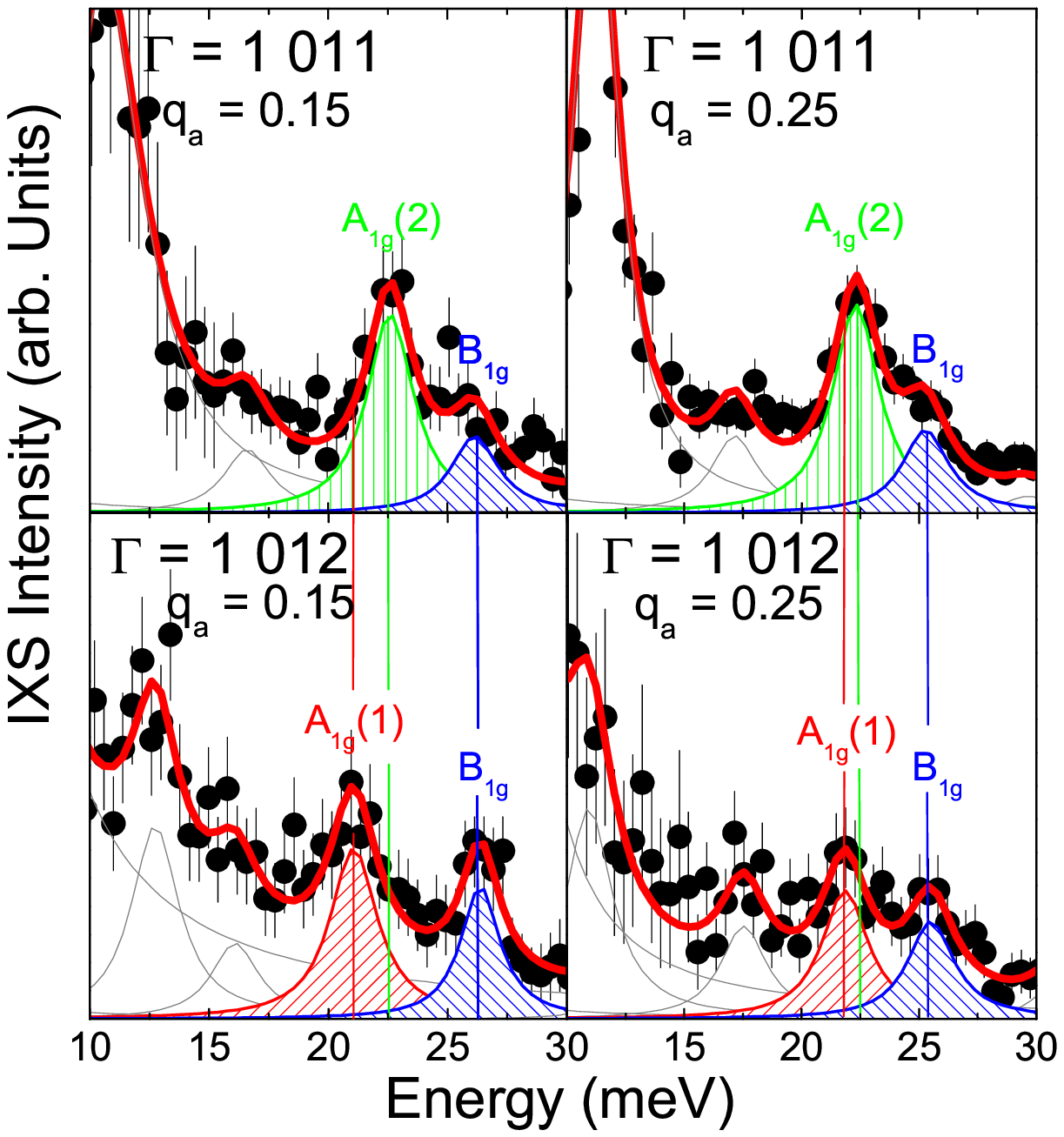}
\end{center}\vspace{-5mm}
\caption{(Color online) Experimental IXS spectra of SmFeAsO$_{0.60}$F$_{0.35}$ for \textbf{q} =(0.15~0~0.17) (left) and \textbf{q} =(0.25~0~0.17) (right) in two different Brillouin zones: $\Gamma$ = (1~0~11) (upper panel) and $\Gamma$ = (1~0~12) (lower panel). The red line is the result of the total fit, and the thin gray lines the individual phonons. The three phonon lines around 23 meV have been highlighted.}
\label{figcontrast}
\end{figure}

The spectra were analyzed by fitting to a series of Lorentzians, as shown in Figs.~\ref{figdoping} and ~\ref{figcontrast}. With the exception of the elastic line, fitting of the data to resolution limited peaks gave rather poor results, and slightly broader lineshapes (2.2 to 2.5 meV) were used. Similar broadening has also been reported in the case of CaFe$_2$As$_2$~\cite{Mittal_PRL2009}. Such broadening is potentially interesting, but widths could not be extracted with enough confidence, due to the low statistics. We thus focus here on the energy of the phonons, which are extracted reliably.

To allow unambiguous assignment of the three modes around 23 meV, we mapped out their dispersion. Only two are visible at small $q_a$ for measurements carried out close to the (1 0 12) $\Gamma$ point, so we also recorded several spectra close to the (1 0 11) $\Gamma$ point for the SmFeAsO$_{0.60}$F$_{0.35}$ sample. This is illustrated in Fig. ~\ref{figcontrast}, where we show energy scans at $q_a$ = 0.15 and $q_a$ = 0.25 in the two different zones [$\Gamma$ = (1 0 12) and $\Gamma$ = (1 0 11)].
The 26 meV mode is present in both zones, while the mode at 23 meV [21 meV] is seen only for measurements carried out from $\Gamma$ = (1 0 11) [$\Gamma$ = (1 0 12)].

\begin{figure}[t!]
\begin{center}
\includegraphics[width=0.95\columnwidth]{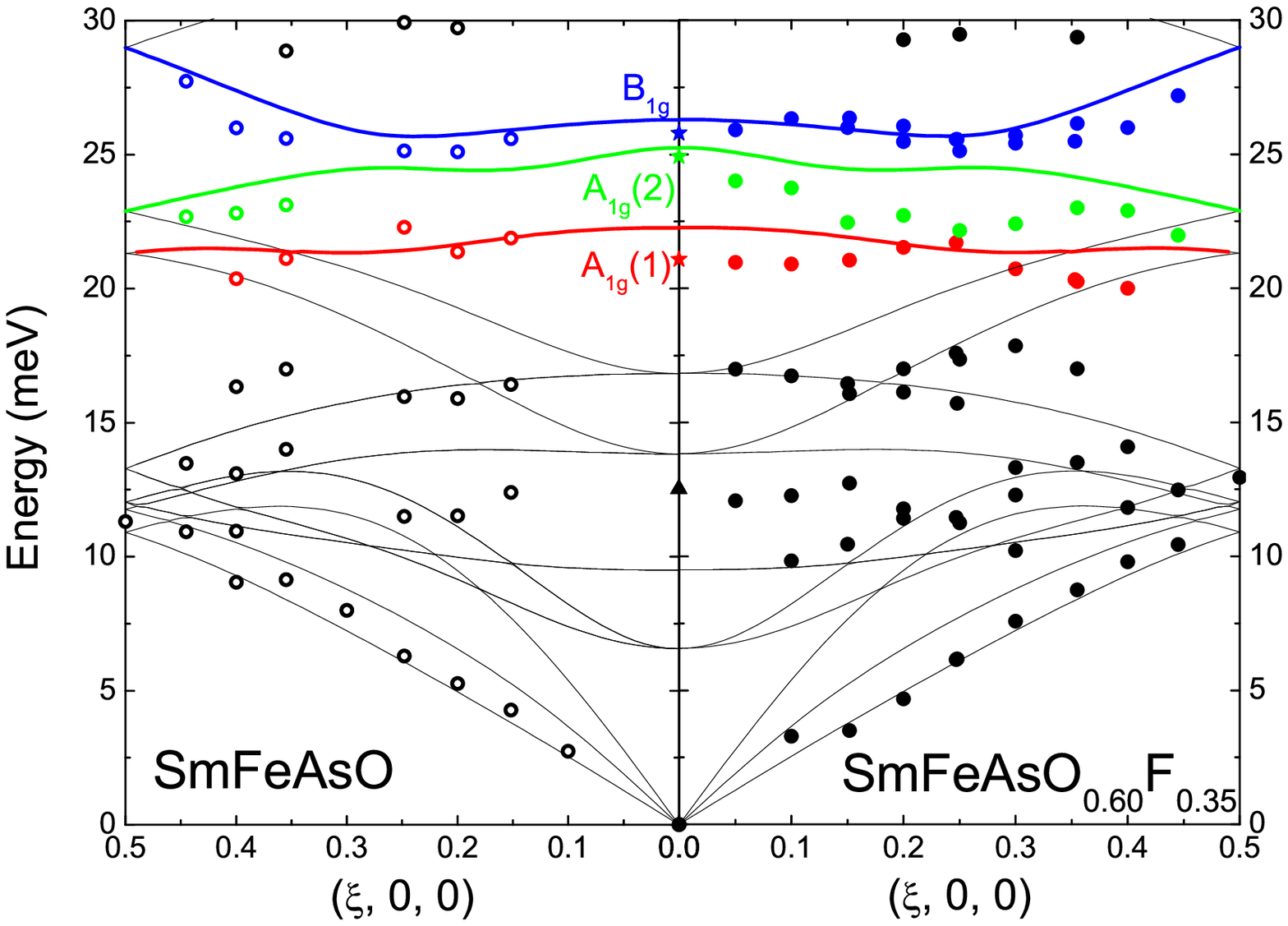}
\end{center}\vspace{-5mm}
\caption{(Color online) Experimental phonon dispersion along the ($q_a$ 0 0) direction of the parent SmFeAsO (left panel) and superconducting SmFeAsO$_{0.60}$F$_{0.35}$ (right panel), at room temperature. Data points with $|q_c| \le 0.17$ where included (The ``doubling" of some points at a given $q_a$ is due to a slightly different $q_c$ in the two measurements). Error bars derived from the fitting are smaller than the data points.
Full lines correspond to the theoretical dispersion of LaFeAsO from ref.~\cite{Noffsinger_PRL2009} and may be considered as a guide to the eye. The three c-axis polarized Raman active modes (see text) have been highlighted. Note that further modes extend up to 55 meV in these materials (not shown).}
\label{figdisp}
\end{figure}

The dispersion of these modes along the ($q_a$~0~0) direction for the two compounds is summarized in Fig.~\ref{figdisp}. The theoretical dispersion for LaFeAsO from Ref.~\cite{Noffsinger_PRL2009} is also shown, and we have added the Raman (stars) and infra-red (IR, triangle) data obtained on SmFeAsO in Refs.~\cite{Hadjiev_PRB2008, Marini_EPL2008}.
Our data extrapolate well to these zone center measurements, allowing us to assign our experimental points to the three c-axis polarized Raman modes: the 21 meV mode is assigned to the in-phase A$_{1g}$ (As,Sm) mode (red, labeled A$_{1g}$(1) in the figures), the 23 meV mode to the out-of-phase A$_{1g}$ (As,Sm) mode (green, A$_{1g}$(2)) and the 26 meV mode to the B$_{1g}$ (Fe,O) mode (blue).
In addition to the transverse acoustic branch at low energy, we note the presence of other branches, one around 12 meV that could be assigned to the IR-active c-axis polarized (Sm,Fe,As) A$_{2u}$ mode, and another around 17 meV, whose energy seems to correspond to the in-plane polarized E$_g$ mode.
However, their precise identifications remains challenging due to the large discrepancies between the IXS intensities predicted by the DFT calculations and the experimental spectra.
Some of these modes also seem to display a doping dependence, and we also note that their relative intensities are strongly renormalized, indicating additional doping-induced changes of the lattice dynamics. A complete study of these phenomena is beyond the scope of this paper, and requires further investigation.

\begin{figure}[!t]
\begin{center}
\includegraphics[width=1\columnwidth]{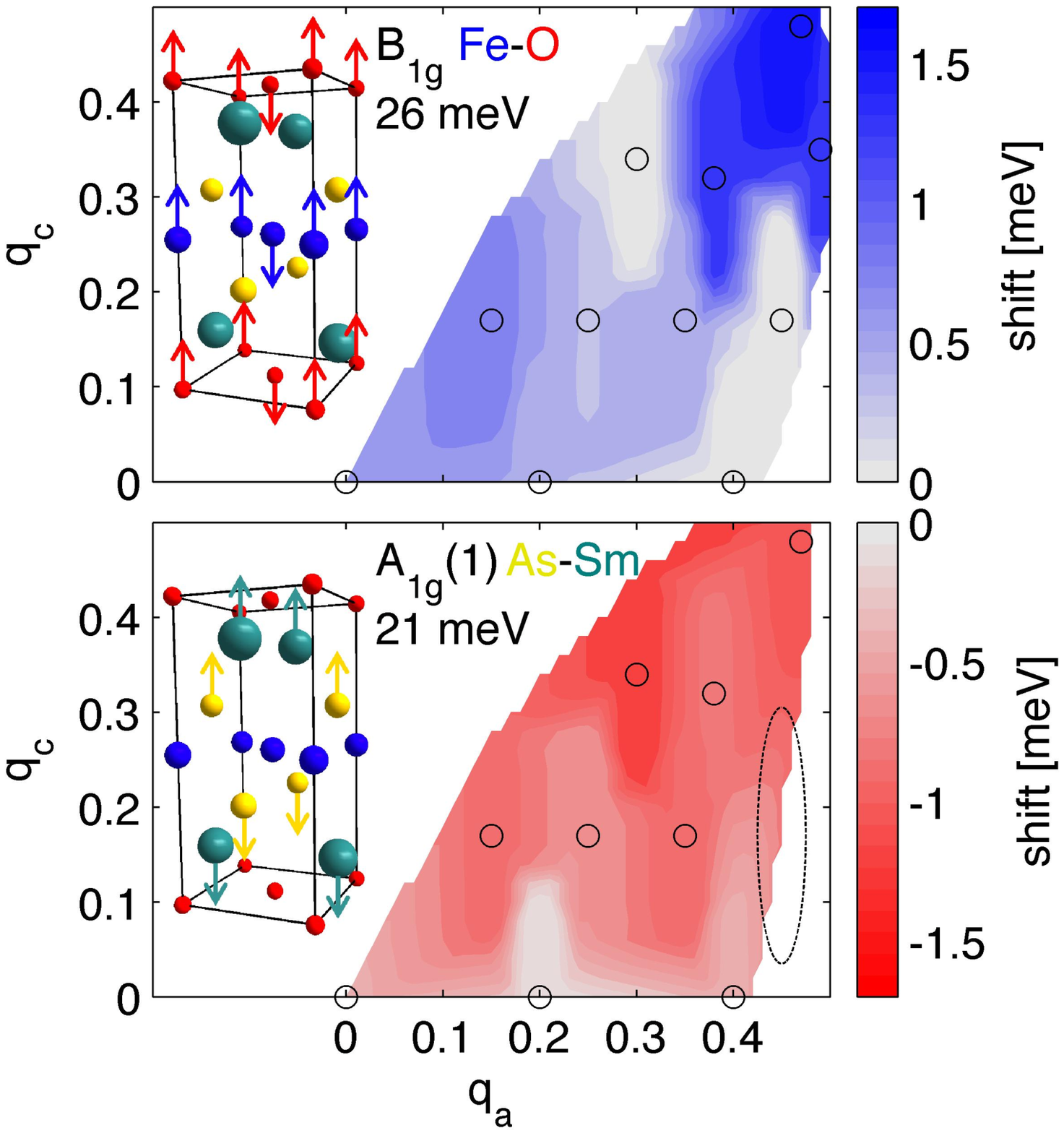}
\end{center}\vspace{-5mm}
\caption{(Color online) Momentum dependence of the doping induced renormalization of the 26 meV,  B$_{1g}$ (upper panel) and 21 meV, in-phase A$_{1g}$ (lower panel) modes in the ($q_a$~0~$q_c$) plane. The empty circles correspond to the analyzer positions, the dashed ellipse indicates the q-resolution, and the white areas where there is insufficient data. The insets show schematics of the respective \bf{q}=0 eigenvectors, with the atoms color coded as indicated.}
\label{figmaps}
\end{figure}

From this assignment and the comparison of doped and undoped spectra shown in Fig.~\ref{figdoping}, it follows that the two modes that display the strongest doping dependence are the in-phase A$_{1g}$ (As,Sm) mode at 21 meV and the B$_{1g}$ (Fe,O) mode at 26 meV.
We have exploited the 8 analyzers available on ID28 to map the \textbf{q}-dependence of the amplitude of the doping-induced shifts of these modes in the ($q_a~0~q_c$) plane. The results are plotted in Fig.~\ref{figmaps}. Remarkably, we find that the doping-induced renormalization of these modes varies strongly with momentum transfer.
The maximum amplitude of the softening of the in-phase A$_{1g}$ mode in the ($q_a~0~q_c$) plane is found to be 1.2 meV close to \textbf{q}=(0.3 0 0.3), while the largest hardening of the B$_{1g}$ mode is found to be 1.7 meV close to \textbf{q}=(0.5 0 0.5).
We note that this doping dependence of the Fe vibration contrasts with recent nuclear resonant IXS measurements~\cite{Higashitaniguchi_PRB2008} which showed no doping dependence of the Fe partial PDOS of LaFeAsO$_{1-x}$F$_x$.

\par
It is worth stressing that our results are not consistent with extant first principle calculations.
Early calculations based on the virtual crystal approximation suggested that doping should induce only minor changes in the phonon spectra~\cite{Boeri}. Our previous PDOS data~\cite{LeTacon_PRB2008} together with the present single crystal data demonstrate the failure of such an approach to properly describe the effects of doping.
More recently, F-doping was explicitly taken into account in a 2x2x1 supercell within an LDA approach~\cite{Noffsinger_PRL2009}. The authors concluded that the doping-induced changes seen in the PDOS~\cite{LeTacon_PRB2008} were strictly related to structural relaxation.
Doping was also predicted to induce hardening for the out-of-phase A$_{1g}$ (As,Sm) and B$_{1g}$ (Fe,O) modes, and a splitting of the in-phase (As,Sm) A$_{1g}$ mode.
Experimentally, we observe a hardening of the B$_{1g}$ mode, but see no evidence for the in-phase A$_{1g}$ mode splitting -- each feature of the undoped sample's spectra has a one-to-one correspondence with a feature observed in the doped compound. Moreover, as emphasized earlier, only a softening of this mode is observed. Finally, for the few points where we have been able to measure the out-of-phase A$_{1g}$ phonon in both samples, no energy renormalization, or only a small softening is observed.

The improved agreement between experiments carried out above $T_N$ in parent 122 compounds and lattice dynamics calculations following the inclusion of spin-polarization~\cite{Hahn_PRB2009} indicates that magnetism is the missing ingredient in the {\em ab-initio} calculations discussed above. In this regard, theoretical studies suggest that the magnetic ground state is intimately related to the Fe-As distance along the c-axis~\cite{Yin_PRL2008, Yildirim_PhysicaC2009}. A corollary to this is that atomic motions modulating this distance, as the in-phase out-of-plane (As,Sm) mode and the (Fe,O) mode do, must couple to the amplitude of the Fe moments~\cite{Yndurain_PRB2009}.
In contrast, the out-of-phase A$_{1g}$ (As,Sm) mode is expected to only weakly change the FeAs tetrahedra and therefore to exhibit only weak spin-phonon coupling~\cite{Yildirim_PhysicaC2009}. Indeed, this phonon shows little or no renormalization which suggests that the effects observed here reflect a coupling of the electronic system to the lattice through this spin channel. Further, coupling may be possible with the AF fluctuations known to be present in the paramagnetic phase~\cite{Kitagawa_JPSP2009, Osborn_PhysicaC2009}.

Our data provide two important clues as to the form of this coupling. First, the opposite shift of the two magnetically-active phonons is highly unusual and suggests that there may be some kind of resonance in the electronic system around 23 meV, with the bare phonons lying on either side of this resonance ~\cite{res-note1}. Second, the maps of the magnitude of the renormalization (Fig.~\ref{figmaps}) provide the first hints as to the momentum dependence of this coupling.
In principle, the momentum and energy dependence of the observed renormalization could have two sources, the electron-phonon coupling, $g(\textbf{q},\omega)$, and the (magnetic or electronic) susceptibility, $\chi(\textbf{q},\omega)$.
Both may evolve with doping, although in the absence of additional theoretical and experimental input it is difficult to resolve their respective contributions to the effects reported here (though, we note that changes in the magnetic susceptibility induced by modifications of the nesting conditions with doping have been discussed~\cite{Yaresko_PRB2009}).
Nevertheless, it is interesting to speculate that the evolution of this coupling, the signatures of which we have observed here in the phonon spectra, may potentially play a role in the dramatic changes in the low temperature ground states.

In summary, we report a strong doping dependence of the two c-axis polarized (As,Sm) A$_{1g}$ and (Fe,O) B$_{1g}$ modes in the paramagnetic state of the SmFeAs(O$_{1-x}$F$_{y}$) iron-pnictide. The doping-induced renormalizations of these two modes have opposite signs and show an unexpected momentum dependence that remains to be understood.
Our results strongly suggest that phonons play an active role in the physics of iron-pnictides and that both further theoretical and experimental investigations are urgently required to clarify this point. Of most relevance here, we note that it has been argued that the coupling between the in-phase A$_{1g}$ mode and the magnetic iron sublattice could (strongly) enhance the electron-phonon coupling in the iron-pnictides~\cite{Yndurain_PRB2009}.

This project was supported by the U.S. Department of Energy, Division of Materials Science, under Contract No. DE-AC02-98CH10886,  by the Swiss National Science Foundation (NCCR MaNEP), the Royal Society, and EPSRC. The authors are grateful to J. v. d. Brink, A. Yaresko and D. Inosov  for useful conversations, and to J. Noffsinger for providing us his DFT calculation.


\begin{thebibliography}{9}
\bibitem{Takahashi_Nature2008} H. Takahashi, K. Igawa, K. Arii, Y. Kamihara, M. Hirano, and H. Hosono, Nature, \textbf{453}, 376 (2008).
\bibitem{Ren_CPL2008} Z.-A. Ren, W. Lu, J. Yang, W. Yi, X.-L. Shen, C. Zheng, G.-C. Che, X.-L. Dong, L.-L. Sun, F. Zhou, and Z.-X. Zhao, Chinese Physics Letters, 2215 (2008).
\bibitem{Boeri} L. Boeri, O. V. Dolgov, and A. A. Golubov, Phys. Rev. Lett., \textbf{101}, 026403 (2008).
\bibitem{Liu_Nature2009} R. H. Liu, T. Wu, G. Wu, H. Chen, X. F. Wang, Y. L. Xie, J. J. Ying, Y. J. Yan, Q. J. Li, B. C. Shi, W. S. Chu, Z. Y. Wu, and X. H. Chen, Nature, \textbf{459}, 64 (2009).
\bibitem{Mukuda_JPSJ2008} H. Mukuda, N. Terasaki, H. Kinouchi, M. Yashima, Y. Kitaoka, S. Suzuki, S. Miyasaka, S. Tajima, K. Miyazawa, P. Shirage, H. Kito, H. Eisaki, and A. Iyo, Journal of the Physical Society of Japan, \textbf{77}, 093704 (2008).
\bibitem{Yin_PRL2008} Z. P. Yin, S. Lebegue, M. J. Han, B. P. Neal, S. Y. Savrasov, and W. E. Pickett, Physical Review Letters, \textbf{101}, 047001 (2008).
\bibitem{Yildirim_PhysicaC2009} T. Yildirim, Physica (Utrecht) \textbf{469C}, 425 (2009).
\bibitem{Fukuda_JPCS2008} T. Fukuda, A. Q. R. Baron, S.-i. Shamoto, M. Ishikado, H. Nakamura, M. Machida, H. Uchiyama, S. Tsutsui, A. Iyo, H. Kito, J. Mizuki, M. Arai, H. Eisaki, and H. Hosono, Journal of the Physical Society of Japan, \textbf{77}, 103715 (2008).
\bibitem{Hahn_PRB2009} S. E. Hahn, Y. Lee, N. Ni, P. C. Canfield, A. I. Goldman, R. J. McQueeney, B. N. Harmon, A. Alatas, B. M. Leu, E. E. Alp, D. Y. Chung, I. S. Todorov, and M. G. Kanatzidis, Physical Review B, \textbf{79}, 220511 (2009).
\bibitem{Rahlenbeck} M. Rahlenbeck, G. L. Sun, D. L. Sun, C. T. Lin, B. Keimer, and C. Ulrich, Physical Review B, \textbf{80}, 064509 (2009).
\bibitem{Chauviere_PRB2009} L. Chauviere, Y. Gallais, M. Cazayous, A. Sacuto, M. A. Measson, D. Colson, and A. Forget, Physical Review B, \textbf{80}, 094504 (2009).
\bibitem{LeTacon_PRB2008} M. Le Tacon, M. Krisch, A. Bosak, J. W. G. Bos, and S. Margadonna, Physical Review B, \textbf{78}, 140505 (2008).

\bibitem{Yndurain_PRB2009} F. Yndurain and J. M. Soler, Phys. Rev. B, \textbf{79} 134506 (2009).
\bibitem{Zhigadlo_JPCM2008} N. D. Zhigadlo, S. Katrych, Z. Bukowski, S. Weyeneth, R. Puzniak, and J. Karpinski, Journal of Physics: Condensed Matter, 342202 (2008).
\bibitem{Karpinski_PhysicaC2009} J. Karpinski, N. D. Zhigadlo, S. Katrych, Z. Bukowski, P. Moll, S. Weyeneth, H. Keller, R. Puzniak, M. Tortello, D. Daghero, R. Gonnelli, I. Maggio-Aprile, Y. Fasano, Ø. Fischer, K. Rogacki, and B. Batlogg, Physica C: Superconductivity, \textbf{469}, 370 (2009).
\bibitem{Burkel_RPP99} E. Burkel, Rep. Prog. Phys., \textbf{63} 171 (1999).
\bibitem{Mittal_PRL2009} R. Mittal, L. Pintschovius, D. Lamago, R. Heid, K. P. Bohnen, D. Reznik, S. L. Chaplot, Y. Su, N. Kumar, S. K. Dhar, A. Thamizhavel, and T. Brueckel, Physical Review Letters, \textbf{102}, 217001 (2009).
\bibitem{Noffsinger_PRL2009} J. Noffsinger, F. Giustino, S. G. Louie, and M. L. Cohen, Physical Review Letters, \textbf{102}, 147003 (2009).
\bibitem{Hadjiev_PRB2008} V. G. Hadjiev, M. N. Iliev, K. Sasmal, Y. Y. Sun, and C. W. Chu, Physical Review B, \textbf{77}, 220505 (2008).
\bibitem{Marini_EPL2008} C. Marini, C. Mirri, G. Profeta, S. Lupi, D. Di Castro, R. Sopracase, P. Postorino, P. Calvani, A. Perucchi, S. Massidda, G. M. Tropeano, M. Putti, A. Martinelli, A. Palenzona and P. Dore, EPL \textbf{84}, 67013 (2008).
\bibitem{Higashitaniguchi_PRB2008} S. Higashitaniguchi, M. Seto, S. Kitao, Y. Kobayashi, M. Saito, R. Masuda, T. Mitsui, Y. Yoda, Y. Kamihara, M. Hirano, and H. Hosono, Physical Review B, \textbf{78}, 174507 (2008).
\bibitem{Kitagawa_JPSP2009} K. Kitagawa, N. Katayama, K. Ohgushi, and M. Takigawa, Journal of the Physical Society of Japan, \textbf{78}, 063706 (2009).
\bibitem{Osborn_PhysicaC2009} R. Osborn, S. Rosenkranz, E. A. Goremychkin, and A. D. Christianson, Physica C: Superconductivity, \textbf{469}, 498 (2009).
\bibitem{res-note1} We note that, intriguingly, scaling the resonance energy observed in the 122 compounds at 15 meV by T$_C$ yields 23 meV.
\bibitem{Yaresko_PRB2009} A. N. Yaresko, G. Q. Liu, V. N. Antonov, and O. K. Andersen, Physical Review B, \textbf{79}, 144421 (2009).
\end{thebibliography}
\end{document}